\documentclass[%
superscriptaddress,
preprint,
 amsmath,amssymb,
 aps,
]{revtex4-2}

\usepackage{graphicx}
\usepackage{dcolumn}
\usepackage{tikz}
\usetikzlibrary{automata,arrows,positioning,calc}
\usepackage{amsthm}
\newtheorem{theorem}{Theorem}

\usepackage{color}

\DeclareMathOperator*{\argmin}{arg\,min}

\begin{document}


\title{Minimal informational requirements for fitness}

\author{Alexander S. Moffett}
\email{asmoffett@cse.yorku.ca}
\affiliation{Department of Electrical Engineering and Computer Science, York University, Toronto, Ontario, Canada}
\author{Andrew W. Eckford}
\email{aeckford@yorku.ca}
\affiliation{Department of Electrical Engineering and Computer Science, York University, Toronto, Ontario, Canada}

\date{\today}

\begin{abstract}
The existing concept of the ``fitness value of information'' provides a theoretical upper bound on the fitness advantage of using information concerning a fluctuating environment. Using concepts from rate-distortion theory, we develop a theoretical framework to answer a different pair of questions: What is the minimal amount of information needed for a population to achieve a certain growth rate? What is the minimal amount of information gain needed for one sub-population to achieve a certain average selection coefficient over another? We introduce a correspondence between fitness and distortion and solve for the rate-distortion functions of several systems using analytical and numerical methods. Because accurate information processing is energetically costly, our approach provides a theoretical basis for understanding evolutionary ``design principles'' underlying information-cost trade-offs.
\end{abstract}

\maketitle

\section{Introduction}\label{sec:introduction}

The ``fitness value of information'' \cite{donaldson2010fitness,rivoire2011value}, addresses the theoretical maximum fitness gain of a population due to information. More specifically, the fitness value of information relates the maximal increase in the growth rate of a population due to information about the environment to information theoretic quantities. The key quantity for measuring the population growth rate, and defining fitness, is the Lyapunov exponent
\begin{equation}\label{eq:lyapunov}
\Lambda\equiv\lim_{t\rightarrow\infty}\frac{1}{t}\log{\frac{N_{t}}{N_{0}}}
\end{equation}
where $N_{t}$ is the population size at time $t$. This work has applied the results of Kelly \cite{kelly1956new}, who related the maximal increase in wealth accumulation rate in gambling on horse races due to side information with the mutual information between the side information and the winning horses. Subsequent work has further developed the Kelly framework as applied to evolution \cite{rivoire2016informations,xue2018benefits,xue2019environment,tal2020adaptive,moffett2020fitness}.

Work on the fitness value of information has largely focused on optimal phenotype strategies given fixed models of environmental cues, which provide information about the true environmental state, and systems for sensing the state of environmental cues. However, the systems through which individuals sense their environment, whether a biochemical signaling network in a bacterium or the nervous system of an animal, are heritable and thus subject to evolutionary processes. To our knowledge, trade-offs between the reliability of information provided by a sensory channel and the metabolic costs of accurate environmental sensing have not been explored in depth. 

In this article, we address two questions distinct from the questions underlying the fitness value of information framework. What is the minimal amount of information about the environment needed for a population to achieve a given growth rate? What is the minimal amount of information gain needed for a sub-population to achieve a gain in average relative fitness over another sub-population? 

We borrow concepts from rate-distortion theory \cite{cover-book}, originally developed to establish theoretical bounds on lossy data compression. We establish 
an equivalence between the fitness of a population, given by Eq. \ref{eq:lyapunov}, and distortion, where distortion in this context corresponds to the mean loss in growth rate due to noisy environmental sensing. The rate-distortion function then describes optimal trade-offs between distortion and information in a growing population. Under reasonable assumptions about the rate-distortion and cost functions, we show that an overall growth rate function accounting for metabolic costs can be defined with a unique optimal distortion and rate given a defined model of the environment. Finally, we indicate how a selection coefficient between two sub-populations can be defined in terms of the rate-distortion and cost functions. Our use of rate-distortion theory to find minimal information requirements for fitness in biological populations adds to the growing number of biological applications of rate-distortion theory \cite{tlusty2008rate,marzen2016weak,iglesias2016use,marzen2017evolution}. 

\section{Rate distortion theory}\label{sec:rd_theory}

We begin by setting up our model of population growth. We consider two competing ``genotypes'', which we label as $A$ and $B$, which constitute sub-populations of the total population. This section concerns the behavior of a population consisting of single genotype, while we will address competition between genotypes $A$ and $B$ in Section \ref{sec:competition}. The individuals of any genotype can divide themselves between phenotypes $\hat{x}\in\hat{\mathcal{X}}$ based on noisy sensing of the environmental state $x\in\mathcal{X}$. This division depends on the strategy $\pi(\hat{x}|y)$, where $y\in\mathcal{Y}$ is the internal representation of the environment within individuals. The strategy describes the probability that a phenotype $\hat{x}$ will be expressed given that an individual perceives the environment as $y$. We consider environmental cues $c\in\mathcal{C}$ available to all individuals chosen according to the cue distribution $q_{\text{env}}(c|x)$, and internal representations chosen according to the sensing distribution $q_{\text{in}}(y|c)$. The environmental cues represent observable aspects of the environment which are informative about the environmental state while the internal representations result from sensory mechanisms within individual organisms in the population \cite{rivoire2011value}. We restrict the sets $\mathcal{X},\mathcal{C},\mathcal{Y},\hat{\mathcal{X}}$ to be finite here. The environment ($X$), environmental cue ($C$), internal representation ($Y$), and phenotype ($\hat{X}$) thus follow a Markov chain
\begin{equation}\label{eq:markov}
X\rightarrow{}C\rightarrow{}Y\rightarrow{}\hat{X}.
\end{equation}
The growth rate of individuals depends on the phenotype expressed as well as the environmental state, and is written as
\begin{equation}
w(x,\hat{x}):\mathcal{X}\times\hat{\mathcal{X}}\rightarrow{}\mathbb{R}^{0+}  
\end{equation}
where $\mathbb{R}^{0+}$ denotes the non-negative real numbers.

For our model, we can easily write the Lyapunov exponent for a sub-population in discrete time as follows \cite{rivoire2011value}
\begin{equation}\label{eq:lyapunov}
\Lambda[q_{\text{in}},q_{\text{env}},\pi]=\sum_{x\in\mathcal{X}}\sum_{c\in\mathcal{C}}p(x)q_{\text{env}}(c|x)\log\Bigg(\sum_{\hat{x}\in\hat{\mathcal{X}}}\sum_{y\in\mathcal{Y}}w(x,\hat{x})\pi(\hat{x}|y)q_{\text{in}}(y|c)\Bigg)
\end{equation}
under the assumption that environmental states are i.i.d. over time steps and that environmental cues, internal representations, and phenotypes are i.i.d. over time steps conditional on environmental states, environmental cues, and internal representations, respectively. Additionally, internal representations and phenotypes are i.i.d. over individual organisms in the population. We write the Lyapunov exponent for a sub-population given environmental cue distribution $q_{\text{env}}(c|x)$ and sensing distribution $q_{\text{in}}(y|c)$ and then optimized over strategy $\pi(\hat{x}|y)$ as
\begin{equation}\label{eq:opt_lyapunov}
\Lambda^{\star}[q_{\text{env}},q_{\text{in}}]\equiv\max_{\pi}\Lambda[q_{\text{env}},q_{\text{in}},\pi].   
\end{equation}
This growth rate is maximized when sensing and cues are noiseless
\begin{equation}\label{eq:opt_lyapunov_max}
\Lambda^{\star}[q_{\text{env}},q_{\text{in}}]\leq{}\Lambda^{\star}[q_\text{env},\delta_{y,c}]\leq{}\Lambda^{\star}[\delta_{c,x},\delta_{y,c}]
\end{equation}
where $\delta_{y,c}$ and $\delta_{c,x}$ are Kronecker delta functions indicating noiseless internal representations and environmental cues, respectively. In order for information about the environment to be useful for growth, and thus for the inequalities in Eq. \ref{eq:opt_lyapunov_max} to be strict, there must be trade-offs between phenotypes. That is, for the cases of interest in this work, there should not be a single phenotype with fitness greater than or equal to the fitness of all other phenotypes in all environments.

Using Eqs. \ref{eq:opt_lyapunov} \& \ref{eq:opt_lyapunov_max}, we define the non-negative quantity
\begin{equation}\label{eq:unopt_cost_of_imperfect_information}
\Omega[q_{\text{env}},q_{\text{in}},\pi]\equiv\Lambda^{\star}[q_{\text{env}},\delta_{y,c}]-\Lambda[q_{\text{env}},q_{\text{in}},\pi]
\end{equation}
and its minimized counterpart
\begin{equation}\label{eq:cost_of_imperfect_information}
\Omega^{\star}[q_{\text{env}},q_{\text{in}}]\equiv\Lambda^{\star}[q_{\text{env}},\delta_{y,c}]-\Lambda^{\star}[q_{\text{env}},q_{\text{in}}]
\end{equation}
which is a modified version of the fitness cost of imperfect sensing \cite{rivoire2011value} depending on both the sensing distribution $q_{\text{in}}(y|c)$ and the cue distribution $q_{\text{env}}(c|x)$. In Eq. \ref{eq:cost_of_imperfect_information}, we compare optimal Lyapunov exponents of perfect and imperfect sensing given the same distribution of environmental cues, in contrast to the fitness cost of imperfect information. The loss in growth rate due to noisy sensing in Eq. \ref{eq:cost_of_imperfect_information} serves as the mean distortion in our rate-distortion framework, as explained below. Because of our definition of mean distortion in Eq. \ref{eq:cost_of_imperfect_information}, we use ``distortion'' and ``growth rate loss'' interchangeably.

The mutual information between environmental states and internal representations
\begin{align}\label{eq:mi_xy}
I(X;Y)&=\sum_{x\in\mathcal{X}}\sum_{y\in\mathcal{Y}}p(x)\bigg(\sum_{c\in\mathcal{C}}q_{\text{in}}(y|c)q_{\text{env}}(c|x)\bigg)\\\nonumber
\times&\log\Bigg(\frac{\sum_{c\in\mathcal{C}}q_{\text{in}}(y|c)q_{\text{env}}(c|x)}{\sum_{x\in\mathcal{X}}p(x)\sum_{c\in\mathcal{C}}q_{\text{in}}(y|c)q_{\text{env}}(c|x)}\Bigg)
\end{align}
describes how informative sensory mechanisms are about the environment on average. When $q_{\text{in}}(y|c)=\delta_{y,c}$, $I(X;Y)=I(X;C)$ while if additionally $q_{\text{env}}(c|x)=\delta_{c,x}$, then $I(X;Y)=H(X)$.

The fitness cost associated with achieving a certain growth rate (or equivalently, loss in growth rate as compared with the maximum possible growth rate), however, only depends on how reliable the sensory mechanism of an organism is, as reflected in Eq. \ref{eq:cost_of_imperfect_information}. This is captured by the mutual information between environmental cues and internal representations
\begin{align}\label{eq:mi_cy}
I(C;Y)&=\sum_{x\in\mathcal{X}}\sum_{c\in\mathcal{C}}\sum_{y\in\mathcal{Y}}p(x)q_{\text{env}}(c|x)q_{\text{in}}(y|c)\\\nonumber
\times&\log\Bigg(\frac{q_{\text{in}}(y|c)}{\sum_{x\in\mathcal{X}}\sum_{c\in\mathcal{C}}p(x)q_{\text{env}}(c|x)q_{\text{in}}(y|c)}\Bigg).
\end{align}
where, from Eq. \ref{eq:markov}, we have $I(C;Y)\geq{}I(X;Y)$ due to the data processing inequality.

We can now define the rate-distortion function as
\begin{equation}\label{eq:rate-distortion}
R(D)=\min_{q_{\text{in}}(y|c):\Omega^{\star}[q_{\text{env}},q_{\text{in}}]\leq{}D}I(C;Y)
\end{equation}
which is the minimal amount of mutual information between the internal representation of the environment and environmental cues necessary needed to achieve a mean distortion of at most $D$. The modified fitness cost of imperfect information $\Omega^{\star}[q_{\text{env}},q_{\text{in}}]$ (Eq. \ref{eq:cost_of_imperfect_information}) functions as the mean distortion, representing the mean growth rate lost by noisy sensing of the environmental cue. Zero distortion is only achievable when perfect information about environmental cues are available to a population, so that $R(0)=H(C)$. On the other hand, the smallest $D$ for which $R(D)=0$ is equal to $\Omega^{\star}[q_{\text{env}},\eta]=I(X;C)$.

We emphasize that in contrast to the ``fitness value of information'' literature, we are interested in optimization over the sensing distribution $q_{\text{in}}(y|c)$ rather than the phenotype expression strategy $\pi(\hat{x}|y)$. In order to define the mean distortion, Eq. \ref{eq:unopt_cost_of_imperfect_information} must be minimized over $\pi(\hat{x}|y)$. This defines a constraint for optimization over $q_{\text{in}}(y|c)$. Because the rate-distortion function in Eq. \ref{eq:rate-distortion} describes optimal trade-offs between growth rate loss and information, the strategy $\pi(\hat{x}|y)$ on the rate-distortion curve is always optimal since otherwise the growth rate loss (distortion) could be reduced using the same amount of information, or the information could be reduced keeping the growth rate loss constant. Overall, this problem is an example of bilevel optimization (see Section \ref{sec:bilevel_opt}).

The distortion function corresponding to the mean distortion (Eq. \ref{eq:cost_of_imperfect_information}) is
\begin{align}\label{eq:distortion_function}
d(x,c&;q_{\text{env}},q_{\text{in}})=\\\nonumber
&\log\bigg(\sum_{\hat{x},y}w(x,\hat{x})\pi^{\star}_{q_{\text{env}},\delta}(\hat{x}|y)\delta_{y,c}\bigg)\\\nonumber
&-\log\bigg(\sum_{\hat{x},y}w(x,\hat{x})\pi^{\star}_{q_{\text{env}},q_{\text{in}}}(\hat{x}|y)q_{\text{in}}(y|c)\bigg)\\\nonumber
=&-\log\bigg(\frac{\sum_{\hat{x},y}w(x,\hat{x})\pi^{\star}_{q_{\text{env}},q_{\text{in}}}(\hat{x}|y)q_{\text{in}}(y|c)}{\sum_{\hat{x},y}w(x,\hat{x})\pi^{\star}_{q_{\text{env}},\delta}(\hat{x}|c)}\bigg)
\end{align}
where $\pi^{\star}_{q_{\text{env}},\delta}(\hat{x}|y)$ is the strategy that maximizes $\Lambda[q_{\text{env}},\delta_{y,c},\pi]$, when internal sensing is noiseless, and $\pi^{\star}_{q_{\text{env}},q_{\text{in}}}(\hat{x}|y)$ is the strategy that maximizes $\Lambda[q_{\text{env}},q_{\text{in}},\pi]$ when internal sensing is distributed according to $q_{\text{in}}(y|c)$. This is a generalized form of the logarithmic loss function \cite{erkip1998efficiency,courtade2013multiterminal,moffett2021code}. We can see that $\Omega^{\star}[q_{\text{env}},q_{\text{in}}]$ is the mean distortion using Eqs. \ref{eq:lyapunov}, \ref{eq:opt_lyapunov}, \ref{eq:cost_of_imperfect_information}, \& \ref{eq:distortion_function} 
\begin{align}
\sum_{x,c}p(x,c)d(x,c;q_{\text{env}},q_{\text{in}})=&\sum_{x,c}p(x)q_{\text{env}}(c|x)\bigg[\log\bigg(\sum_{\hat{x},y}w(x,\hat{x})\pi^{\star}_{q_{\text{env}},\delta}(\hat{x}|y)\delta_{y,c}\bigg)\\
&-\log\bigg(\sum_{\hat{x},y}w(x,\hat{x})\pi^{\star}_{q_{\text{env}},q_{\text{in}}}(\hat{x}|y)q_{\text{in}}(y|c)\bigg)\bigg]\\
=&~\Lambda^{\star}[q_{\text{env}},\delta_{y,c}]-\Lambda^{\star}[q_{\text{env}},q_{\text{in}}]=\Omega^{\star}[q_{\text{env}},q_{\text{in}}].
\end{align}
The ``test channel'' consists purely of $q_{\text{in}}(y|c)$, representing the ability of organisms to evolve their sensory capabilities, while we assume that the reliability of the external environmental cue cannot be influenced by evolutionary processes. An equivalent rate-growth function can be defined
\begin{equation}\label{eq:rate-growth}
R(W)=\min_{q_{\text{in}}(y|c):\Lambda^{\star}[q_{\text{env}},q_{\text{in}}]\geq{}W}I(C;Y)
\end{equation}
which is the minimal amount of mutual information needed between the environmental cue and internal representation needed to achieve an average growth rate of at least $W$.

The mean distortion is bounded by the generalized entropy defined by Rivoire and Leibler \cite{rivoire2011value} minus the entropy of the environmental state given the environmental cue
\begin{align}\label{eq:distortion_bound}
&\Omega^{\star}[q_{\text{env}},q_{\text{in}}]\leq{}H_{p}^{(q_{\text{env}},q_{\text{in}})}-H(X|C)\\\nonumber
&=-\sum_{x,c}p(x)q_{\text{env}}(c|x)\log\Bigg(\frac{p(c)\sum_{y}\pi_{q_{\text{env}},q_{\text{in}}}^{\star}(x|y)q_{\text{in}}(y|c)}{q_{\text{env}}(c|x)p(x)}\Bigg)\\\nonumber
&=D_{\text{KL}}\Big(q_{\text{env}}(c|x)p(x)~\Big|\Big|~\sum_{y}\pi_{q_{\text{env}},q_{\text{in}}}^{\star}(x|y)q_{\text{in}}(y|c)p(c)\Big)
\end{align}
with equality when $w(x,\hat{x})>0$ only when $x=\hat{x}$, the marginal probability of environmental cue $c$ is $p(c)=\sum_{x}q_{\text{env}}(c|x)p(x)$, and $D_{\text{KL}}(\cdot~||~\cdot)$ is the Kullback-Leibler divergence. For most cases, it is difficult to obtain $R(D)$ explicitly, although we will find an exact formula for $R(D)$ in a several simple cases.

\section{Interpretation of the rate-distortion function}\label{sec:interpretation}

\begin{figure}
    \centering
    \includegraphics[width=.8\textwidth]{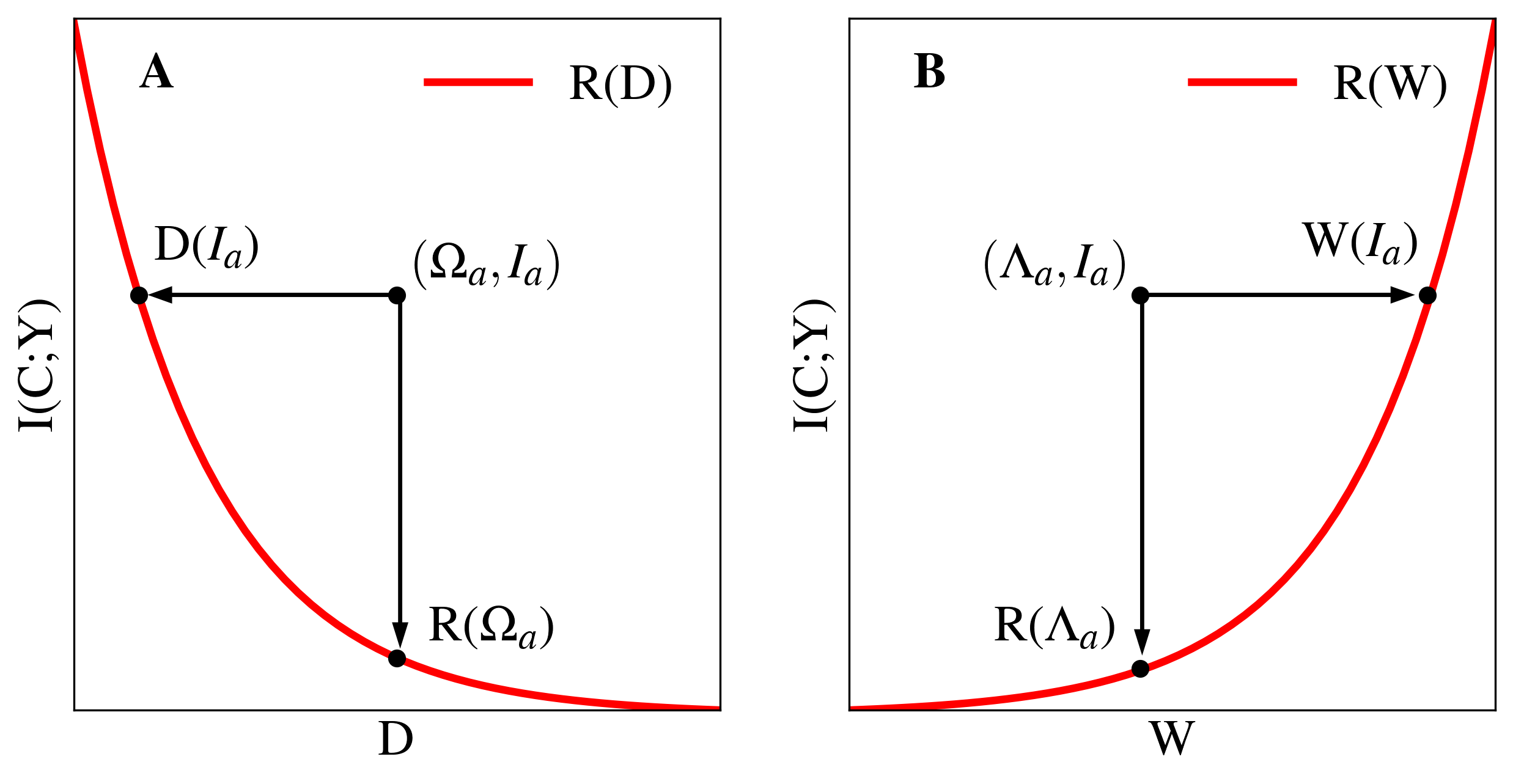}
    \caption{A) The rate-distortion function, $R(D)$ describes optimal sensory systems and strategies. The point $(\Omega_{a},I_{a})$ is not optimal, as it lies above $R(D)$. The organisms of this genotype could change their strategies, given an $I_{a}$-achieving sensory system, to reach a distortion of $D(I_{a})$, where $D(R)$ is the inverse the rate-distortion function. Although $I_{a}$ remains constant, the sensory system itself can change to achieve the optimal $q_{\text{in}}(y|c)$ such that $I(C;Y)=I_{a}$. The population could also evolve a sensory system that achieves $R(\Omega_{a})$, adopting the corresponding optimal strategy. Of course, arbitrary $(\Omega_{b},I_{b})$ could be achieved, provided that $I_{b}\geq{}R(\Omega_{b})$. B) The rate-growth function, which can be described similarly to the rate-distortion function except that instead of mean distortion $\Omega$, we are concerned with the long-term growth rate $\Lambda$.}
    \label{fig:rate-distortion_diagram}
\end{figure}

Each pair of $\Omega^{\star}[q_{\text{env}},q_{\text{in}}]$ and $I(C;Y)$, which we will call $(\Omega,I)$ pairs, represents the loss in growth rate of the population due to noisy sensing and the mutual information between environmental cues and the internal cues of a population. The rate distortion function $R(D)$ represents the minimum mutual information necessary to achieve distortion of at most $D$. Alternatively, the inverse function $D(R)$ represents the minimum distortion achievable with mutual information $R$. Then $(\Omega,I)$ pairs lying above the rate-distortion curve are sub-optimal, in that the metabolic costs paid to achieve $R$ could be used to achieve a lower distortion, $D(R)$, while the distortion $D$ could be achieved with less information, $R(D)$. On the other hand, $(\Omega,I)$ pairs lying below the rate-distortion curve are unachievable. The rate-distortion function thus represents optimal trade-offs between growth rate loss and information processing ability, and can be viewed as a Pareto front \cite{shoval2012evolutionary} between the two. 

Non-optimal $(\Omega,I)$ pairs can occur in several ways. Consider a distortion rate pair $(\Omega_{a},I_{a})$ with $I_{a}>R(\Omega_{a})$ (Fig. \ref{fig:rate-distortion_diagram}A). For a fixed $q_{\text{in}}(y|c)$, if a sub-optimal strategy $\pi_{q_{\text{env}},q_{\text{in}}}(\hat{x}|y)$ is used, then the distortion $\Omega_{a}$ will be larger than $D(I_{a})$, where $I_{a}$ is determined by $p(x)$, $q_{\text{env}}(c|x)$ and $q_{\text{in}}(y|c)$. On the other hand, depending on the probability distributions $p(x)$ and $q_{\text{env}}(c|x)$, there can exist more than one distribution $q_{\text{in}}(y|c)$ that yield the same distortion $\Omega_{a}$ but multiple values of $I(C;Y)$, with some greater than $R(\Omega_{a})$, including $I_{a}$. The rate $R(\Omega_{a})$ could be achieved through an optimal choice of both sensing distribution $q_{\text{in}}(y|c)$ and strategy $\pi(\hat{x}|y)$. Thus, optimal $(\Omega,I)$ pairs are only possible with both optimal strategies and with the choice of $q_{\text{in}}(y|c)$ that satisfies Eq. \ref{eq:rate-distortion}. We can consider the equivalent framework of the rate-growth function (Fig. \ref{fig:rate-distortion_diagram}B) analogously.

Because $q_{\text{in}}(y|c)$ reflects the information processing abilities of an organism, which is subject to evolution, the rate-distortion function represents an optimal ``design principle'' for sensory systems. We emphasize that evolutionary processes will not necessarily drive sensory systems towards the rate-distortion function. Because accurate sensing of environmental conditions is metabolically costly down to the level of biochemical signaling \cite{mehta2012energetic,lang2014thermodynamics,sartori2014thermodynamic,mehta2016landauer}, there should be metabolic costs associated with $I(C;Y)$ which will decrease the resources available for growth.

\section{The rate-distortion function and bilevel optimization}\label{sec:bilevel_opt}

In contrast to more conventional rate-distortion problems, the distribution over which we optimize actually changes the distortion function itself (Eq. \ref{eq:distortion_function}), rather than the distribution of $x$ and $c$. The optimization problem defined in Eq. \ref{eq:rate-distortion} is an example of bilevel optimization \cite{sinha2017review}, and can be reformulated as such:
\begin{align}
\min_{q_{\text{in}}}~I(C;Y)\quad\quad\quad&\\\nonumber
\text{subject to:}\quad\quad&\\
\Omega[q_{\text{env}},q_{\text{in}},\pi]&\leq{}D\\
\pi&\in\argmin_{\pi}~\Omega[q_{\text{env}},q_{\text{in}},\pi]\\
q_{\text{in}}(y|c),\pi(\hat{x}|y)&\geq{}0~\forall{}c\in\mathcal{C},\forall{}y\in\mathcal{Y},\forall{}\hat{x}\in\hat{\mathcal{X}}\\
\sum_{y\in\mathcal{Y}}q_{\text{in}}(y|c)&=1~\forall{}c\in\mathcal{C}\\
\sum_{x\in\hat{\mathcal{X}}}\pi(\hat{x}|y)&=1~\forall{}y\in\mathcal{Y}.
\end{align}
In general, these problems are difficult to solve. However, there are a number of methods that have found previous success \cite{sinha2017review}, including evolutionary optimization algorithms. While we do not explore these solution methods in this work, these approaches could be useful for solving more complicated cases than we consider here. 

\section{Explicit costs of sensing}\label{sec:costs}

Up to this point we have not explicitly dealt with metabolic costs to sensing. Here we assume that the cost of sensing $\Gamma{}:\mathbb{R}^{0+}\rightarrow{}\mathbb{R}^{0+}$ is a monotonically increasing function of the sensing mutual information. We write the cost as $\Gamma[I(C;Y)]$ or as a composition $\Gamma\circ{}R(D)$ when optimal $(\Omega,I)$ pairs on the rate-distortion function are achieved. We also assume that $\Gamma[0]=0$, that is, the metabolic cost of informationless sensory mechanisms is zero. The overall growth rate when considering costs is then
\begin{equation}\label{eq:overall_growth_not_optimal}
G[q_{\text{env}},q_{\text{in}},\pi]=\Lambda^{\star}[q_{\text{env}},\delta_{y,c}]-\Omega[q_{\text{env}},q_{\text{in}},\pi]-\Gamma[I(C;Y)].
\end{equation}
When the rate-distortion function is achieved, the overall optimal growth rate as a function of $D$ is 
\begin{equation}\label{eq:overall_growth}
G^{\star}(D)=\Lambda^{\star}[q_{\text{env}},\delta_{y,c}]-D-\Gamma\circ{}R(D) 
\end{equation}
so that $G[q_{\text{env}},q_{\text{in}},\pi]\leq{}G^{\star}(\Omega[q_{\text{env}},q_{\text{in}},\pi])$. To find the optimal distortion $\hat{D}$ when the rate-distortion function is achieved, where $G^{\star}(D)<G^{\star}(\hat{D})$ for all $D\neq{}\hat{D}$, we take the derivative of $G^{\star}(D)$ with respect to $D$ and set it equal to zero to yield
\begin{equation}
\frac{d}{dD}\Gamma\circ{}R(D)\Big|_{D=\hat{D}}=-1.
\end{equation}
If $R(D)$ is continuous, strictly convex, twice continuously differentiable, and monotonically decreasing in $D$ over $\big[0,\Omega^{\star}[q_{\text{env}},\eta]\big]$ and $\Gamma[I(C;Y)]$ is continuous, convex, twice continuously differentiable, and monotonically decreasing in $I(C;Y)$ for $I(C;Y)\geq{}0$, then $G^{\star}(D)$ is concave in $D\in\big[0,\Omega^{\star}[q_{\text{env}},\eta]\big]$. Then there is either a unique critical point $\hat{D}$ on $\big[0,\Omega^{\star}[q_{\text{env}},\eta]\big]$ that maximizes $G^{\star}(\hat{D})$ over $D\geq{}0$ or there is no such critical point and either $\hat{D}=0$ or $\hat{D}=\Omega^{\star}[q_{\text{env}},\eta]$ maximizes $G^{\star}(D)$ over $D\geq{}0$ (see Appendix \ref{appendix:gopt} for proof).

This means that under reasonable assumptions for the rate-distortion function in many cases and for the cost function, there will be a single optimal distortion $\hat{D}$. A genotype that achieves a distortion-rate pair equal to $(\hat{D},R(\hat{D}))$ will have a growth rate $G^{\star}(\hat{D})$ greater than or equal to the growth rate of any other possible genotype. This also means that $R(\hat{D})$ is the amount of information needed about the environment so that no other genotype can achieve a larger overall growth rate. If the optimal strategies and sensory systems are achievable, any sensory system with $I(C;Y)>R(\hat{D})$ is sub-optimal, incurring larger costs than the benefit of the information.

A simple class of cost functions that we will explore in later sections is
\begin{equation}
\Gamma[I(C;Y)]=\alpha{}I(C;Y)^{n}
\end{equation}
where $n$ is either $1$ or a positive, even integer and $\alpha>0$. All cost functions in this class are convex and monotonically increasing for $0\leq{}I(C;Y)\leq{}H(C)$. Monotonically increasing, strictly convex cost functions correspond to diminishing returns, meaning that at higher mutual information values a given increase in information will be more costly. That is, these conditions mean that for mutual information values $I_{A}$ and $I_{B}$ and an increase in mutual information $\Delta{}I$
\begin{equation}
I_{A}<I_{B}\implies\Gamma[I_{A}+\Delta{}I]-\Gamma[I_{A}]<\Gamma[I_{B}+\Delta{}I]-\Gamma[I_{B}].
\end{equation}

\section{Example: Binary environment with sensing through a symmetric channel}\label{sec:example_bsc} 

In the first set of examples we consider, we assume that sensing occurs through a binary symmetric channel with error probability $\epsilon$. We start with the extremes cases of informationless and noiseless environmental cues. While the case of informationless cues is trivial, these two simple systems are analytically tractable and provide theoretical, if not biological, insight. The section concludes with the more general case of intermediate environmental cue reliability.

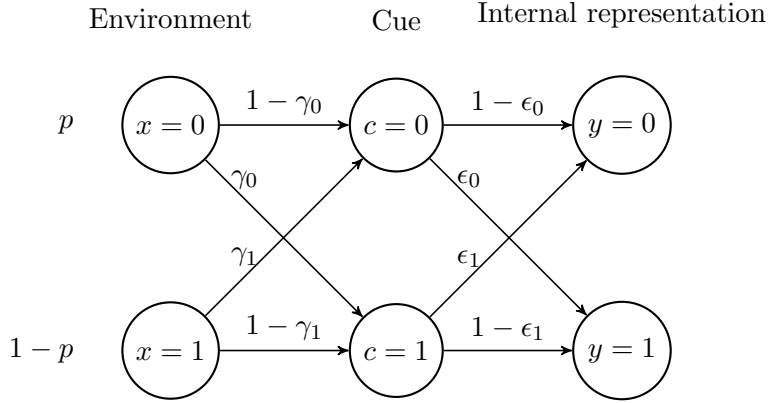
\begin{figure}
\begin{center}
\begin{tikzpicture}[->, >=stealth', auto, semithick, node distance=3cm]
\tikzstyle{every state}=[fill=white,draw=black,thick,text=black,scale=1]
\node[state](x0)             {$x=0$};
\node[](l1)[above=.5 of x0]{Environment};
\node[state](x1)[below of=x0]{$x=1$};
\node[](p0)[left=.5 of x0]{$p$};
\node[](p1)[left=.5 of x1]{$1-p$};
\node[state](c0)[right of=x0]{$c=0$};
\node[](l2)[above=.5 of c0]{Cue};
\node[state](c1)[below of=c0]{$c=1$};
\node[state](y0)[right of=c0]{$y=0$};
\node[](l3)[above=.5 of y0]{Internal representation};
\node[state](y1)[below of=y0]{$y=1$};

\path
(x0) edge node{$1-\gamma_{0}$} (c0)
     edge node[above,pos=.25]{$\gamma_{0}$} (c1)
(x1) edge node[above,pos=.25]{$\gamma_{1}$} (c0)
     edge node{$1-\gamma_{1}$} (c1)
(c0) edge node{$1-\epsilon_{0}$} (y0)
     edge node[above,pos=.25]{$\epsilon_{0}$} (y1)
(c1) edge node[above,pos=.25]{$\epsilon_{1}$} (y0)
     edge node{$1-\epsilon_{1}$} (y1)
;
\end{tikzpicture}
\end{center}
\caption{Diagram of a general binary system with environmental cue error probabilities $\gamma_{0}$ and $\gamma_{1}$ and sensing error probabilities $\epsilon_{0}$ and $\epsilon_{1}$. Environment $x=0$ occurs with probability $p$ and environment $x=1$ occurs with probability $1-p$. In Section \ref{sec:example_bsc}, $\epsilon_{0}=\epsilon_{1}=\epsilon$, while in Section \ref{sec:example_gbc} $\epsilon_{0}$ and $\epsilon_{1}$ can differ.}
\end{figure}

\subsection{Informationless environmental cues}

We now examine how the rate-distortion function and growth rate behave for a simple model of growth. We explore the case where two environmental states, external cues, internal representations, and two phenotypes are possible ($|\mathcal{X}|=|\mathcal{C}|=|\mathcal{Y}|=|\hat{\mathcal{X}}|=2$). We assume that the external cues and internal representations behave as binary symmetric channels (BSCs), so that
\begin{align}
q_{\text{env}}(1|0)&=q_{\text{env}}(0|1)=\gamma\\ q_{\text{in}}(1|0)&=q_{\text{in}}(0|1)=\epsilon 
\end{align}
where $0\leq{}\gamma,\epsilon\leq{}1/2$. Additionally, we initially assume that $w(x,\hat{x})=0$ for $x\neq\hat{x}$ and $w(x,\hat{x})>0$ for $x=\hat{x}$. Now, consider the case where $\gamma=1/2$, meaning that the environmental cue carries no information about the actual state of the environment. The optimal strategy for this case is proportional betting \cite{donaldson2008phenotypic,donaldson2010fitness}, so that
\begin{align}
\pi(0|0)&=\pi(0|1)=p(0)=p\\ 
\pi(1|0)&=\pi(1|1)=p(1)=1-p. 
\end{align}
As the environmental cue is entirely uninformative about the environment, ignoring the state of sensory mechanisms is optimal. Proportional betting leads to a mean distortion of
\begin{equation}
\Omega^{\star}[\eta,q_{\text{in}}]=0
\end{equation}
where $\eta$ indicates that the environmental cue is informationless, which indicates that the mean distortion is unrelated to environmental sensing. Then, the rate-distortion function is
\begin{equation}
R(D)=0,~D\in[0,\infty)
\end{equation}
which indicates that no information is needed to achieve any distortion. Because the only achievable distortion is $0$, it is clear that $\hat{D}=0$ and the maximal overall growth rate is 
\begin{equation}
G^{\star}(\hat{D})=\Lambda^{\star}[\eta,\delta_{y,c}].
\end{equation}
\begin{figure}
    \centering
    \includegraphics[width=\textwidth]{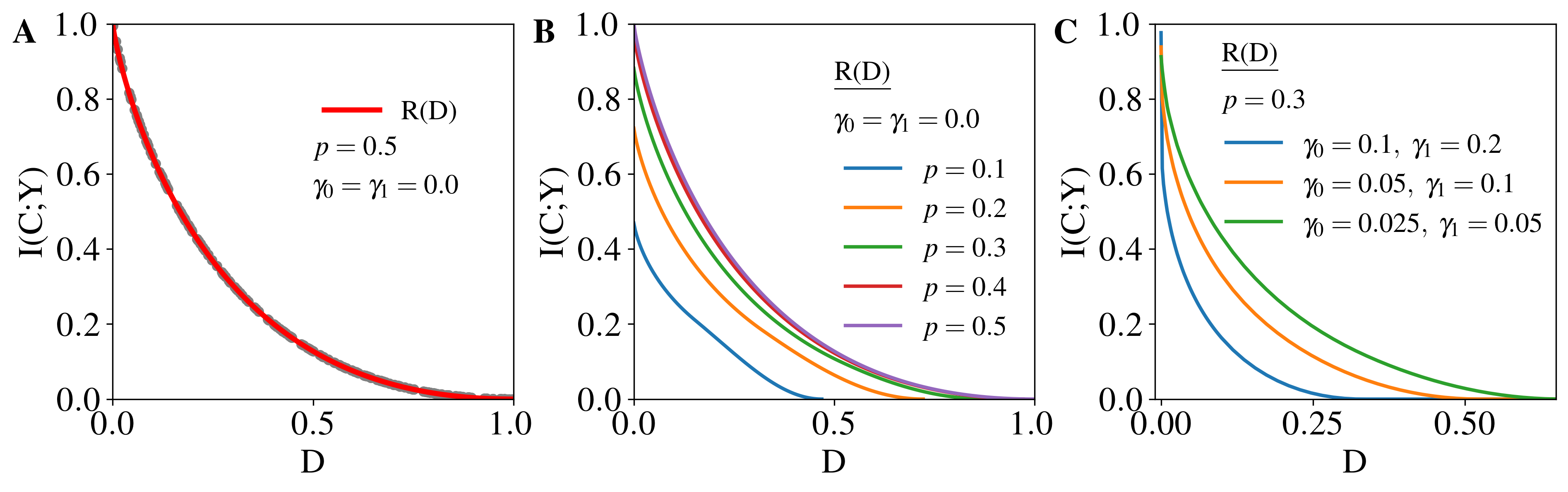}
    \caption{A) The rate-distortion function (red line) for a binary system with $p=1/2$, noiseless environmental cues, and a sensing distribution described by a BSC. The grey dots represent randomly sampled rate-distortion pairs. Note that because of the one-to-one relationship between $\epsilon$ and $D$ in this system, all rate-distortion pairs fall on $R(D)$. B) Rate-distortion functions for the same system with varying values of $p$. As the $p$ deviates from 1/2, less information is required to achieve a given distortion. C) Rate-distortion functions with all parameters fixed except for the environmental cue error probabilities $\gamma_{0}$ and $\gamma_{1}$.}
    \label{fig:fig1}
\end{figure}

\subsection{Noiseless environmental cues}

If instead we assume that environmental cues are completely reliable ($\gamma=0$), the optimal strategy depends on $\epsilon$ in the following manner \cite{rivoire2011value}. If $\epsilon=1/2$, there is as much available information about the environment as when $\gamma=1/2$ and the optimal strategy is again
\begin{align}
\pi(0|0)&=\pi(0|1)=p\\ 
\pi(1|0)&=\pi(1|1)=1-p. 
\end{align}
If $\epsilon=0$, there is no noise in either the environmental cues or the sensing mechanism, reflected by the optimality of ``all in'' strategies
\begin{align}
\pi(0|0)&=\pi(1|1)=1\\
\pi(1|0)&=\pi(0|1)=0.
\end{align}
%
If instead $0<\epsilon<1/2$, at least one of $\pi(0|0)$ and $\pi(1|1)$ will be equal to one in the optimal strategy. Under the assumptions for this model, this leads to a mean distortion of \cite{rivoire2011value}
\begin{equation}\label{eq:bsc_general_distortion}
\Omega^{\star}[\delta_{c,x},q_{\text{in}}]=
\begin{cases}
-\log(1-\epsilon)&\quad\text{for}~\epsilon\leq{}\epsilon_{c}(p)\\
H_{b}(p)-\epsilon_{c}(p)\log\Big(\frac{1-\epsilon}{\epsilon}\Big)&\quad\text{for}~\epsilon>\epsilon_{c}(p)\\
\end{cases}
\end{equation}
where $\epsilon_{c}(p)=\min(p,1-p)$. Because the mean distortion is a monotonically increasing function of $\epsilon$, in the special case where $p=1/2$, we can write $\epsilon$ in terms of $D$ according to
\begin{equation}\label{eq:error_distortion}
\epsilon=1-2^{-D}
\end{equation}
while $I(C;Y)$ is a monotonically decreasing function of $\epsilon$. This means that we can substitute Eq. \ref{eq:error_distortion} into Eq. \ref{eq:mi_cy} to find the rate distortion function. Defining the function
\begin{equation}\label{eq:rate-distortion_explicit_part}
\phi(D)=(1-D)2^{-D}+\big(1-2^{-D}\big)\Big(1+\log\big(1-2^{-D}\big)\Big)
\end{equation}
we can then write the rate distortion function explicitly as
\begin{equation}\label{eq:rate-distortion_explicit}
R(D)=
\begin{cases}
\phi(D)&\quad\text{for}~0\leq{}D\leq{}1\\
0&\quad\text{for}~D>1.\\
\end{cases}
\end{equation}
\begin{figure}
    \centering
    \includegraphics[width=.5\textwidth]{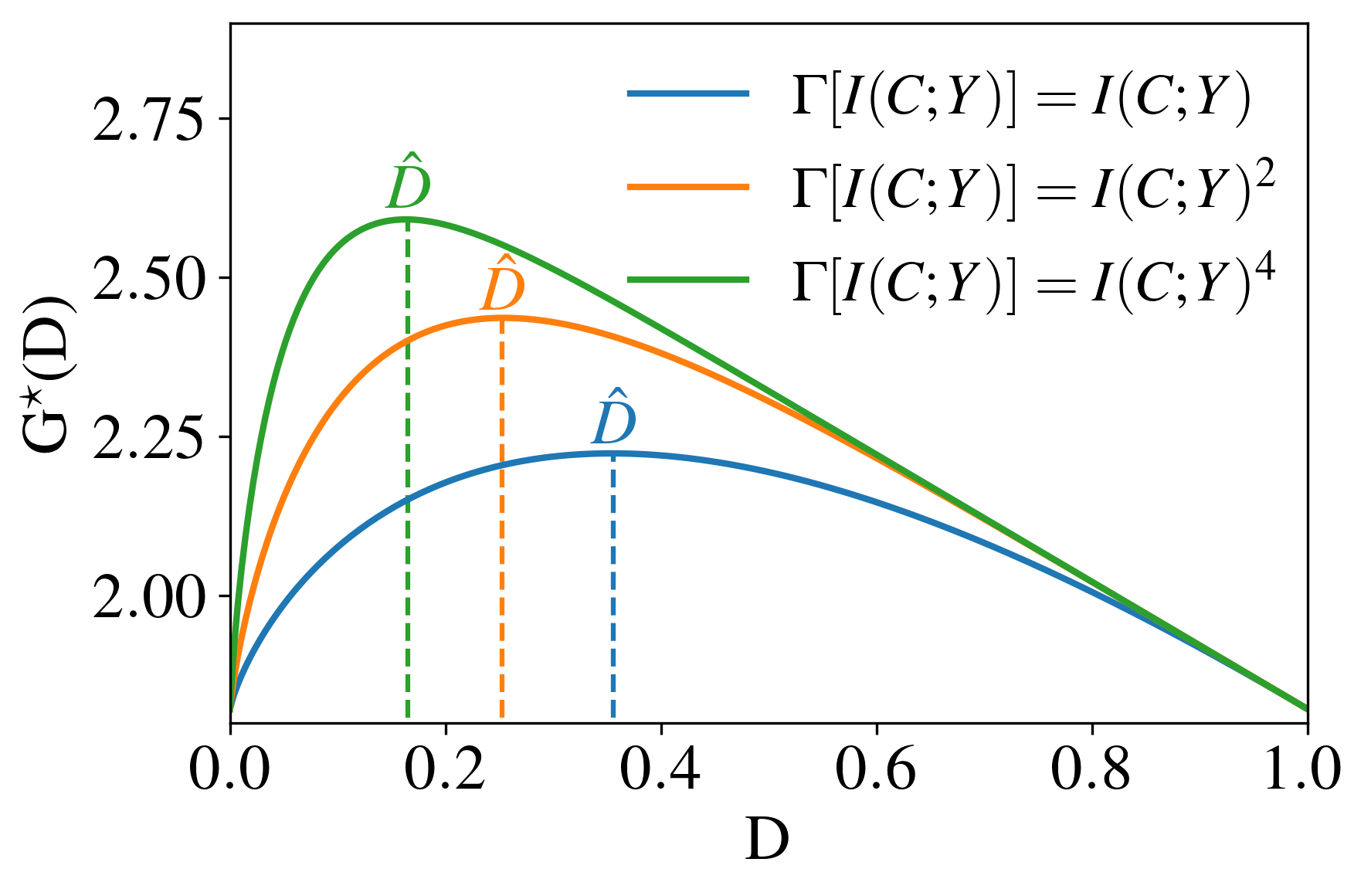}
    \caption{The optimized overall growth function with a BSC sensing channel as a function of distortion for several convex cost functions. The dashed vertical lines indicate the optimal distortion, $\hat{D}$, for each case. Because this is a binary system, $R(D)\leq{}H(X)\leq{}1$. As a result, the cost function $\Gamma\circ{}R(D)=R(D)^{4}$ induces a smaller optimal distortion than $\Gamma\circ{}R(D)=R(D)^{2}$, which in turn induces a smaller optimal distortion than $\Gamma\circ{}R(D)=R(D)$. Because $R(D)$ is monotonically decreasing for $D\in\big[0,\Omega^{\star}[q_{\text{env}},\eta]\big]$, this means that as cost functions grow faster in $R(D)$ it is optimal to have more informative sensory systems. One would usually expect the opposite, that costlier sensing would drive the optimal system towards less accurate sensing. This can indeed be the case when $R(D)>1$ for some values of $D$.}
    \label{fig:fig2}
\end{figure}
The bound on the information needed to achieve a certain mean distortion established by Eq. \ref{eq:rate-distortion_explicit} is demonstrated in Fig. \ref{fig:fig1}A. However, the fact that the individual sensory channel is a BSC means that there is a one-to-one relationship between distortion and rate, so that for \emph{all} $\epsilon$ the bound $R(D)$ is met. If instead $0<p<1/2$, we can define $R(D)$ parametrically through $\epsilon$ according to Eqs. \ref{eq:mi_cy} \& \ref{eq:bsc_general_distortion} as a function of $\epsilon$. Varying $p$ while keeping all other parameters the same as Fig. \ref{fig:fig1}A, we can see in Fig. \ref{fig:fig1}B the effects of altering the probability of each environment. As $|p-1/2|$ increases the entropy of the environment decreases. Fig. \ref{fig:fig1}B shows how more certain environments in this situation require less information to achieve a given distortion. 

\subsection{Noisy environmental cues}

In between the cases of informationless and noiseless environmental cues is the more interesting case of noisy (but not informationless) environmental cues. In this subsection we allow for non-symmetric environmental cue distributions, so that
\begin{align}
q_{\text{env}}(1|0)&=\gamma_{0}\\
q_{\text{env}}(0|1)&=\gamma_{1}\\ q_{\text{in}}(1|0)&=q_{\text{in}}(0|1)=\epsilon.
\end{align}
Fixing $p=0.3$, Fig. \ref{fig:fig1}C shows how varying the reliability of the environmental cue alters the rate-distortion function. As $c$ becomes less reliable (as $\gamma_{0}$ and $\gamma_{1}$ increase) the rate-distortion function reaches the $D$-axis more rapidly with $D$. To achieve the same distortion, less information is required when the environmental cue is less reliable. We can make sense of this seemingly counter-intuitive result by recalling that the term $\Lambda^{\star}[q_{\text{env}},\delta_{y,c}]$ will change with $q_{\text{env}}$. As the environmental cue becomes more reliable, $\Lambda^{\star}[q_{\text{env}},\delta_{y,c}]$ will increase. As a result, a given distortion corresponds to different growth rates if we compare the three cases shown in Fig. \ref{fig:fig1}C. 

\section{Example: Binary environment with sensing through a general channel}\label{sec:example_gbc}

\begin{figure}
    \centering
    \includegraphics[width=.4\columnwidth]{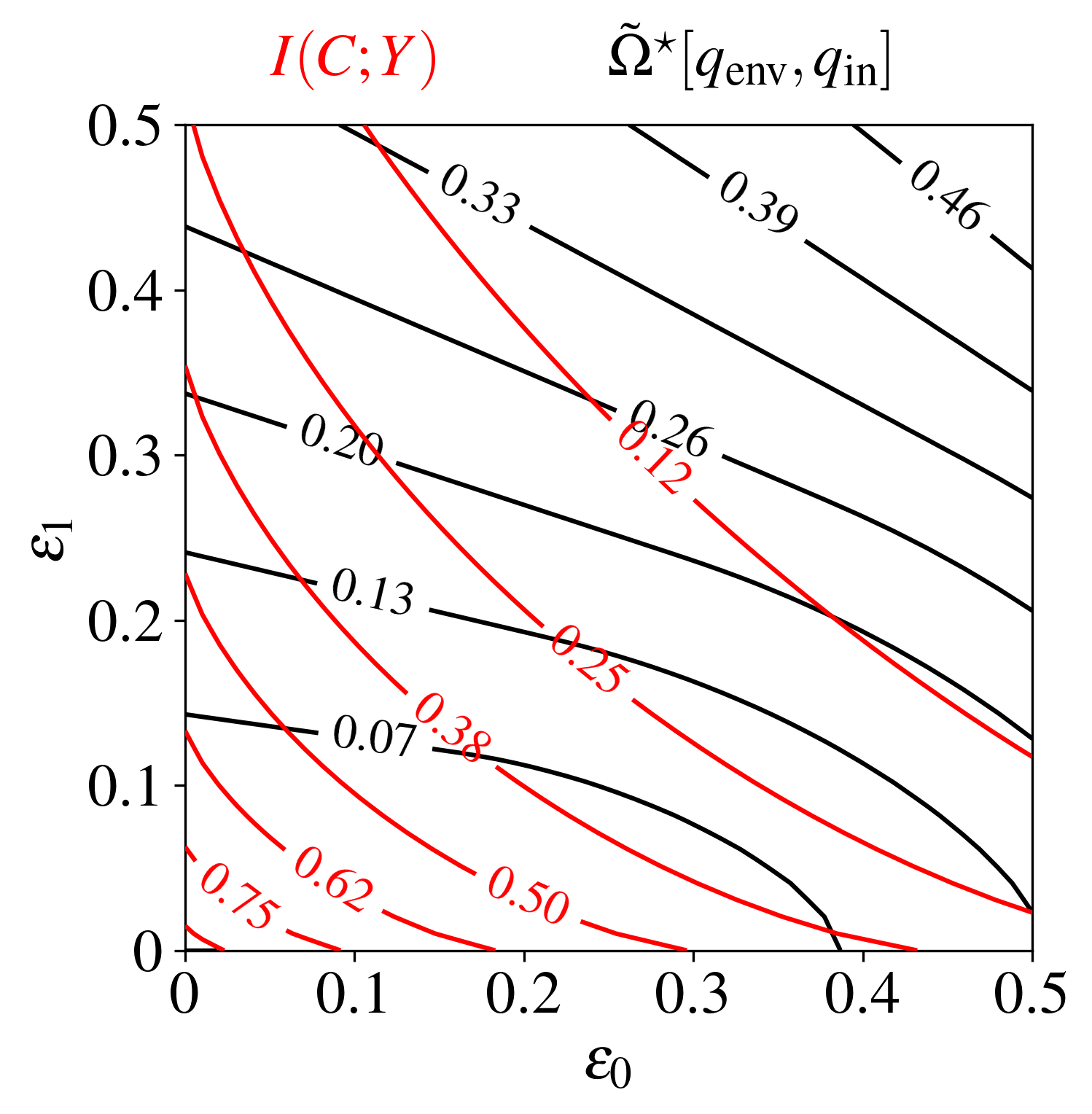}
    \caption{Contour lines of $I(C;Y)$ (shown in red) and $\tilde{\Omega}^{\star}[q_{\text{env}},q_{\text{in}}]$ (shown in black) over the two error probabilities for sensing through a general binary channel. A constraint of the distortion $\tilde{\Omega}^{\star}[q_{\text{env}},q_{\text{in}}]\leq{}D$ corresponds to the region below, and including, the $\tilde{\Omega}^{\star}[q_{\text{env}},q_{\text{in}}]=D$ contour line. Any value of $I(C;Y)$ in this region is then achievable under the constraint. The value of $R(D)$ for some distortion $D$ corresponds to the $I(C;Y)$ contour line that is tangent to the $\tilde{\Omega}^{\star}[q_{\text{env}},q_{\text{in}}]=D$ contour line. The parameters in this figure are $p=0.3$, $\gamma_{0}=0.05$, and $\gamma_{1}=0.1$.}
    \label{fig:figx}
\end{figure}

We now examine the case where sensing of a binary environment occurs through an unconstrained, general binary channel. This model system is of greater relevance to actual biological systems, and we adapt it in Section \ref{sec:example_iabr} to a model of inducible antibiotic resistance in bacteria.

\begin{figure}
    \centering
    \includegraphics[width=.75\columnwidth]{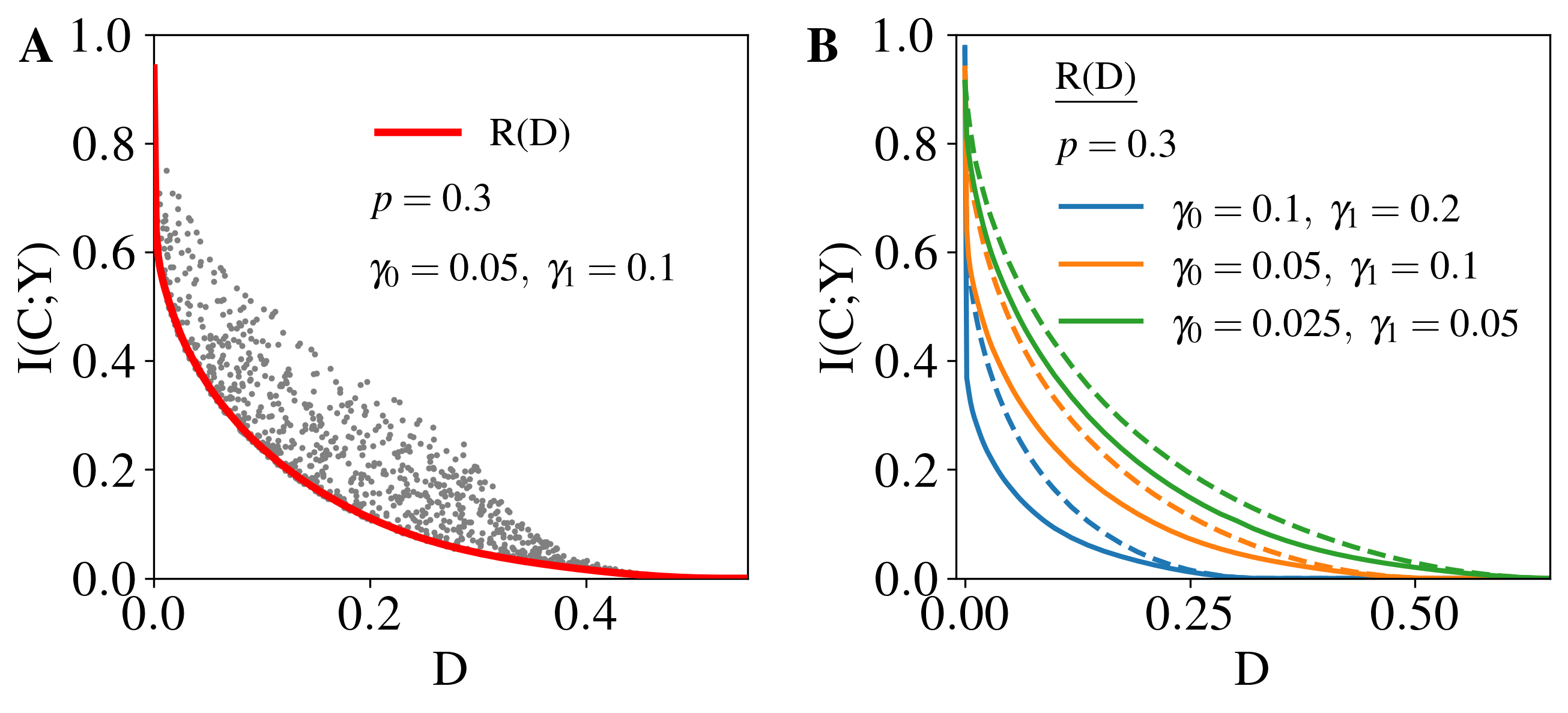}
    \caption{A) The rate-distortion function (red line) for a binary system with noiseless environmental cues and with a sensing distribution described by a general binary channel. The grey dots represent randomly sampled rate-distortion pairs, which are bounded from below by $R(D)$. B) Rate-distortion functions with all parameters fixed except for the environmental cue error probabilities $\gamma_{0}$ and $\gamma_{1}$. The dashed lines indicate the corresponding rate-distortion functions from Fig. \ref{fig:fig2}C where the binary sensory channel was constrained to be symmetric.}
    \label{fig:fig3}
\end{figure}

The error probability in environment $x=0$ ($x=1$) is $\epsilon_{0}$ ($\epsilon_{1}$), and it is possible to find pairs $(\epsilon_{0},\epsilon_{1})\in[0,1/2]\times{}[0,1/2]$ which do not meet the bound set by $R(D)$. In other words, because $q_{\text{in}}(y|c)$ is no longer constrained to be symmetric, it is possible to achieve a fixed distortion $D$ with multiple different sensing distributions (Fig. \ref{fig:fig3}). We also relax the assumption of a diagonal growth matrix (meaning that in this section $x\neq{}\hat{x}$ does not imply that $w(x,\hat{x})=0$) and the symmetry restriction on the environmental cue channel. All together, the environmental cue and sensory distributions can be written as
\begin{align}
q_{\text{env}}(1|0)&=\gamma_{0}\\
q_{\text{env}}(0|1)&=\gamma_{1}\\ q_{\text{in}}(1|0)&=\epsilon_{0}\\
q_{\text{in}}(0|1)&=\epsilon_{1}.
\end{align}
Even in this simple case, it is difficult to explicitly write $\Omega^{\star}[q_{\text{env}},q_{\text{in}}]$ in terms of $p$, $\gamma_{0}$, $\gamma_{1}$, $\epsilon_{0}$, $\epsilon_{1}$, and $w(x,\hat{x})$. Instead, given $p$, $\gamma_{0}$, $\gamma_{1}$, and $w(x,\hat{x})$, we first numerically find $\Omega^{\star}[q_{\text{env}},q_{\text{in}}]$ for a finite number of $(\epsilon_{0},\epsilon_{1})$ pairs and use linear interpolation to find $\Omega^{\star}[q_{\text{env}},q_{\text{in}}]$ for arbitrary $(\epsilon_{0},\epsilon_{1})$ pairs in $[0,1/2]\times{}[0,1/2]$. We refer to this approximate function as $\tilde{\Omega}^{\star}[q_{\text{env}},q_{\text{in}}]$. Next, we numerically minimize $I(C;Y)$ over $\epsilon_{0}$ and $\epsilon_{1}$ using $\tilde{\Omega}^{\star}[q_{\text{env}},q_{\text{in}}]$ in the constraint definitions. This approximation method is used for all results in this section.

To see why it is possible to achieve rate-distortion pairs that do not lie on $R(D)$ when sensing is through a general binary channel, we can overlay the contours of $I(C;Y)$ with those of $\tilde{\Omega}^{\star}[q_{\text{env}},q_{\text{in}}]$, as shown for an example in Fig. \ref{fig:figx} with $I(C;Y)$ contours shown in red and $\tilde{\Omega}^{\star}[q_{\text{env}},q_{\text{in}}]$ in black. With the two independent error probabilities of the general binary channel, we can see that there are many distinct sensing distributions leading to each value of $I(C;Y)$ and $\tilde{\Omega}^{\star}[q_{\text{env}},q_{\text{in}}]$. For example, if we fix $D$ at $0.20$, it is clear that there are $(\epsilon_{0},\epsilon_{1})$ pairs that satisfy $\tilde{\Omega}^{\star}[q_{\text{env}},q_{\text{in}}]\leq{}D=0.20$ and that can achieve some point on each of the displayed $I(C;Y)$ contours. The smallest $I(C;Y)$ achievable with the constraint $\tilde{\Omega}^{\star}[q_{\text{env}},q_{\text{in}}]\leq{}D=0.20$ corresponds to the $I(C;Y)$ contour that is tangent to the $\tilde{\Omega}^{\star}[q_{\text{env}},q_{\text{in}}]=0.2$ contour, which should be close to $I(C;Y)=0.12$. 

We numerically determined the rate-distortion function with parameters $p=0.3$, $\gamma_{0}=0.05$, and $\gamma_{1}=0.1$, shown in Fig. \ref{fig:fig3}A. The estimate of $R(0.20)\approx{}0.12$ from Fig. \ref{fig:figx} (with the same parameters as in Fig. \ref{fig:figx}) matches with the calculation of $R(0.20)$ shown in Fig. \ref{fig:fig3}A. As noted previously, randomly sampled sensory channels for a general binary channel can achieve rate-distortion pairs above the rate-distortion function. These rate-distortion pairs represent organisms that are not fully exploiting the information they are gathering. 

As with the BSC, as the environmental cue becomes less reliable less information is needed to achieve the same distortion (Fig. \ref{fig:fig3}B). Relaxing the constraint of channel symmetry allows for less required information at each distortion between zero and $I(X;C)$ as compared with the BSC.
\begin{figure}
    \centering
    \includegraphics[width=.5\textwidth]{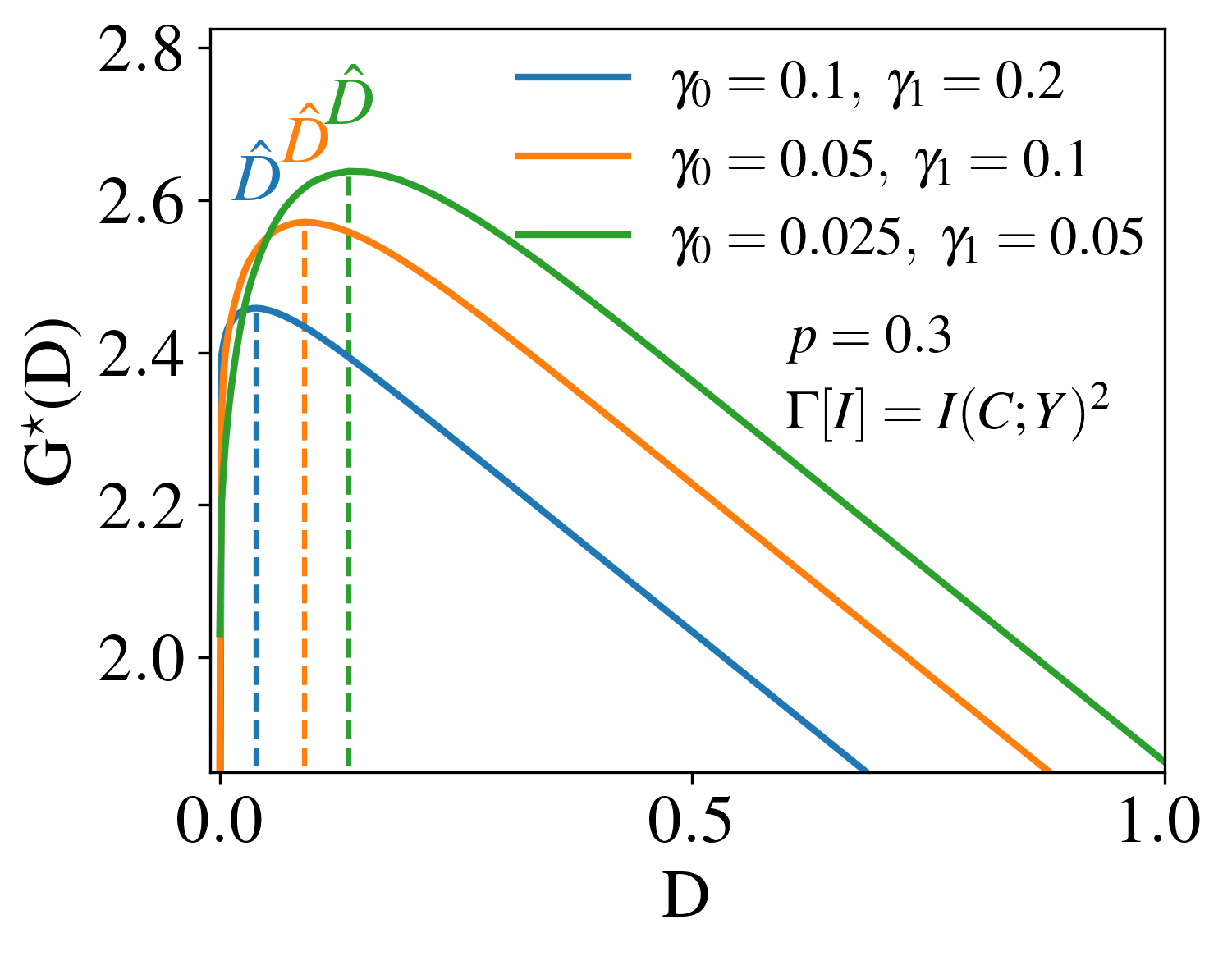}
    \caption{Overall growth rate for the general binary channel with varying environmental cue reliability. As the environmental cue becomes more reliable, both the optimal distortion $\hat{D}$ and the maximal overall growth rate $G^{\star}(\hat{D})$ increase. For the blue growth curve ($\gamma_{0}=0.1,~\gamma_{1}=0.2$) $R(\hat{D})\approx{}0.193$, for the orange growth curve ($\gamma_{0}=0.05,~\gamma_{1}=0.1$) $R(\hat{D})\approx{}0.259$, and for the green growth curve ($\gamma_{0}=0.025,~\gamma_{1}=0.05$) $R(\hat{D})\approx{}0.299$. The cost function for this example is $\Gamma[I(C;Y)]=I(C;Y)^{2}$}
    \label{fig:fig4}
\end{figure}
We can examine how this variation in environmental cue accuracy effects the overall growth rate as a function of $D$. Choosing the cost function $\Gamma[I(C;Y)]=I(C;Y)^{2}$ and setting $p=0.3$, we can see that there is an optimal distortion $\hat{D}$ for each choice of environmental cue distribution (Fig. \ref{fig:fig4}). This optimal distortion increases as environmental cues become more reliable, requiring more reliable sensory mechanisms to achieve $G^{\star}(\hat{D})$ as the rate-distortion functions at $\hat{D}$ are $R(\hat{D})\approx{}0.193$ ($\gamma_{0}=0.1,~\gamma_{1}=0.2$), $R(\hat{D})\approx{}0.259$ ($\gamma_{0}=0.05,~\gamma_{1}=0.1$), and $R(\hat{D})\approx{}0.299$ ($\gamma_{0}=0.025,~\gamma_{1}=0.05$). We note that while more accurate sensory mechanisms are necessary to achieve $G^{\star}(\hat{D})$ for less noisy environmental cues, a genotype $A$ in the presence of a more reliable environmental cue than is available to genotype $B$ can achieve the growth rate $G^{\star}_{B}(\hat{D}_{B})$ at some distortion $D_{A}>\hat{D}_{B}$. In the example shown in Fig. \ref{fig:fig4}, comparing the orange ($\gamma_{0}=0.05,~\gamma_{1}=0.1$) and green ($\gamma_{0}=0.025,~\gamma_{1}=0.05$) curves we find that $G^{\star}_{\text{green}}(0.229)>G^{\star}_{\text{orange}}(\hat{D}_{\text{orange}})$ while $R_{\text{green}}(0.229)<R_{\text{orange}}(\hat{D}_{\text{orange}})$, where $\hat{D}_{\text{orange}}\approx{}0.090$.

\section{Example: inducible antibiotic resistance in bacteria}\label{sec:example_iabr}

Some bacteria have evolved the ability to express genes conferring antibiotic resistance (ABR) only in the presence of a particular antibiotic, reducing fitness costs associated with expression of the ABR genes in the absence of antibiotics \cite{depardieu2007modes}. For example, in \emph{Enterococci} the presence of the antibiotic vancomycin is detected through a two-component signaling system consisting of an integral membrane histidine kinase receptor and a cytoplasmic response regulator protein. Genes conferring ABR are then expressed when the two-component signaling system is activated by antibiotics. If we assume that ABR gene expression is binary (that is, not graded but either off or maximally on), we can adapt the results of Section \ref{sec:example_gbc} to this specific biological example. We consider the environmental state $x=0$ ($x=1$) to be the case where antibiotic concentration is below (above) the threshold concentration. The phenotypic state $\hat{x}=0$ ($\hat{x}=1$) corresponds to when antibiotic resistance genes are not (are) expressed. The fitness function is then 
\begin{align}
w(x=0,\hat{x}=0)&=a\\
w(x=0,\hat{x}=1)&=a-c\\
w(x=1,\hat{x}=0)&=a-d\\
w(x=1,\hat{x}=1)&=a-c-d+b.
\end{align}
The parameter $a$ is a base growth rate when antibiotic concentration is low and resistance genes are not expressed, $c$ is the cost of gene expression, $d$ accounts for the reduction in fitness due to antibiotics, and $b$ is the fitness benefit of expressed antibiotic resistance genes. The parameters defining the fitness function should be ordered according to
\begin{equation}
0 < c < b < d < a,
\end{equation}
meaning that antibiotics cannot reduce the growth rate to zero and that the benefits of ABR gene expression outweigh the costs but do not entirely counteract the fitness effects of antibiotics. 
\begin{figure}
    \centering
    \includegraphics[width=.75\columnwidth]{rd_abr_fig1.png}
    \caption{Rate-distortion and rate-growth functions for the inducible ABR model with varied environmental cue reliability and probability of antibiotic absence. Keeping $p$ constant while varying $\gamma$, the A) rate-distortion functions and B) growth-rate functions show how more reliable environmental cues allows for noisier sensors with a constant distortion, or equivalently growth rate. When the probability of antibiotic absence is varied with a constant environmental cue reliability, the C) rate-distortion and D) rate-growth functions can clearly become non-convex. Increasing $p$ shifts the rate-growth function to larger growth rates. Shared parameters for all subplots are: $a=2$, $b=0.5$, $c=0.25$, and $d=1.25$.}
    \label{fig:fig8}
\end{figure}
The environmental cue distribution in this case corresponds to a Z-channel
\begin{align}
q_{\text{env}}(0|0)&=1\\
q_{\text{env}}(1|0)&=0\\
q_{\text{env}}(0|1)&=\gamma\\
q_{\text{env}}(1|1)&=1-\gamma
\end{align}
considering that the accuracy of concentration sensing is limited due to the stochastic nature of ligand diffusion, binding, and unbinding while no such fundamental difficulty exists for detecting the absence of a ligand. Thus, $\gamma$ represents fundamental limits to concentration sensing (relating to the Berg-Purcell limit \cite{berg1977physics,kaizu2014berg}) which could change depending on factors such as the size of bacterial cells or the antibiotic concentration associated with environmental state $x=1$.

Numerically solving for the rate-distortion function as in Section \ref{sec:example_gbc}, we find as before that increasing the reliability of environmental cues lowers the rate-distortion curve (Fig. \ref{fig:fig8}A). This corresponds to lower informational requirements for achieving a given growth rate with more reliable environmental cues (Fig. \ref{fig:fig8}B). We also varied the probability of antibiotics being absent with fixed environmental cue accuracy, which yielded non-convex rate-distortion functions (Fig. \ref{fig:fig8}C). Viewed instead as rate-growth functions, as it becomes more likely that antibiotics are not present, the rate-growth curve shifts to larger values of $W$ (Fig. \ref{fig:fig8}D).
\begin{figure}
    \centering
    \includegraphics[width=\columnwidth]{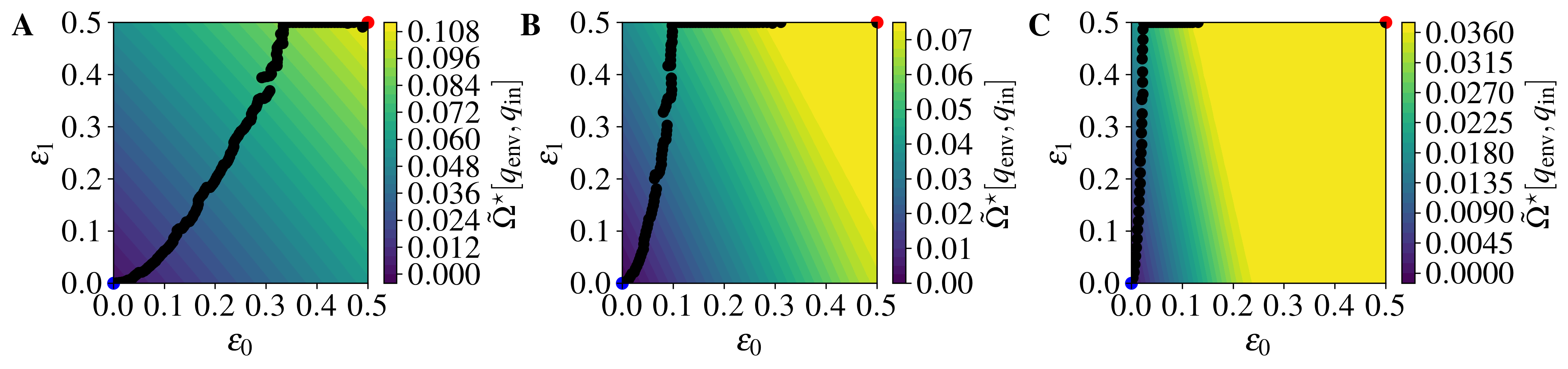}
    \caption{The estimated mean distortion $\tilde{\Omega}^{\star}[q_{\text{env}},q_{\text{in}}]$ as a function of sensing error probabilities $\epsilon_{0}$ and $\epsilon_{1}$ shown with probability of no antibiotic A) $p=0.7$, B) $p=0.8$, and C) $p=0.9$ for the inducible ABR model. The blue filled circle at the origin of each subplot correspond to the optimal sensing distribution when $D=0$ and the red filled circle corresponds to the sensing distribution where $I(C;Y)=0$. In between the black filled circles indicate the estimated optimal error probabilities with varied distortion constraints. Shared parameters for all subplots are: $a=2$, $b=0.5$, $c=0.25$, $d=1.25$, and $\gamma=0.1$.}
    \label{fig:fig9}
\end{figure}
In order to examine the origins of the non-convexity of the rate-distortion functions in Fig. \ref{fig:fig8}C, we can examine the calculated optimal sensing distributions as $D$ increases (Fig. \ref{fig:fig9}). For the apparently convex rate-distortion function with $p=0.7$, we can see that starting from the origin of Fig. \ref{fig:fig9}A and ending at $\epsilon_{0}=\epsilon_{1}=0.5$, there is a gradual trajectory of sensing distributions with increasing distortion. However, for $p=0.8$ (Fig. \ref{fig:fig9}B) and $p=0.9$ (Fig. \ref{fig:fig9}C) there are plateaus in the mean distortion as a function of $\epsilon_{0}$ and $\epsilon_{1}$ where mean distortion is constant. Thus, once this plateau is reached in the trajectory of optimal sensing distributions, any change in the sensing distribution towards the informationless distribution characterized by $\epsilon_{0}=\epsilon_{1}=0.5$ does not change the distortion but reduces the mutual information, $I(C;Y)$. This explains the sudden jumps in the corresponding rate-distortion functions, resulting in their non-convexity.

\section{Comparative advantages between two genotypes}\label{sec:competition}

We now consider the growth of sub-populations corresponding to two genotypes. We assume that the two sub-populations grow independently from one another conditional on the environment and environmental cue. Thus, in our model genotypes do not compete for resources, but achieve different growth rates through sensing the environment. We can write the relative growth rate between $A$ and $B$ as
\begin{align}\label{eq:relative_fitness_general}
\frac{G[q_{\text{env}},q_{\text{in,A}},\pi_{A}]}{G[q_{\text{env}},q_{\text{in,B}},\pi_{B}]}&=\frac{G[q_{\text{env}},q_{\text{in,B}},\pi_{B}]+G[q_{\text{env}},q_{\text{in,A}},\pi_{A}]-G[q_{\text{env}},q_{\text{in,B}},\pi_{B}]}{G[q_{\text{env}},q_{\text{in,B}},\pi_{B}]}=1+s\\
s&\equiv{}\frac{G[q_{\text{env}},q_{\text{in,A}},\pi_{A}]-G[q_{\text{env}},q_{\text{in,B}},\pi_{B}]}{G[q_{\text{env}},q_{\text{in,B}},\pi_{B}]}
\end{align}
where $s$ is an average selection coefficient. All selection coefficients in this section result from the averaging done to define the long-term growth rate in Eq. \ref{eq:lyapunov}, and we emphasize that these quantities cannot be used interchangeably with selection coefficients defined in a single environment \cite{cvijovic2015fate}. In our model, the individual growth rate function $w(x,\hat{x})$ is the same for both genotypes (though this assumption could be relaxed), while changes to the sensor distribution and strategy realized through evolutionary processes differentiate the two genotypes. Using Eq. \ref{eq:overall_growth} we can write the selection coefficient as 
\begin{align}\label{eq:selection_coefficient1}
s&=s_{c}+s_{a}\\\label{eq:selection_coefficient2}
s_{c}&=\frac{\Gamma[I_{B}(C;Y)]-\Gamma[I_{A}(C;Y)]}{G[q_{\text{env}},q_{\text{in,B}},\pi_{B}]}\\\label{eq:selection_coefficient3}
s_{a}&=\frac{\Omega[q_{\text{env}},q_{\text{in,B}},\pi_{B}]-\Omega[q_{\text{env}},q_{\text{in,A}},\pi_{A}]}{G[q_{\text{env}},q_{\text{in,B}},\pi_{B}]}
\end{align}
where $s$ is decomposed into the contributions from metabolic costs ($s_{c}$) and adaptive benefits ($s_{a}$) \cite{lynch2015bioenergetic}. 
In a competitive setting between genotypes $A$ and $B$, it is clear from Eqs. \ref{eq:selection_coefficient1}-\ref{eq:selection_coefficient3} that in order for $A$ to have a larger growth rate than $B$ ($s>0$), the condition
\begin{equation}\label{eq:cost_criterion1}
\Gamma[I_{A}(C;Y)]-\Gamma[I_{B}(C;Y)]<\Omega[q_{\text{env}},q_{\text{in,B}},\pi_{B}]-\Omega[q_{\text{env}},q_{\text{in,A}},\pi_{A}]
\end{equation}
must be met.

Consider a genotype $B$, with distortion-rate pair $(\Omega_{B},I_{B})$. What is the smallest gain in information that another genotype $A$ will need to achieve a selection coefficient of $s$ over genotype $B$? The maximum selection coefficient possible over genotype $B$ is
\begin{equation}
s_{\text{max}}=\frac{G^{\star}(\hat{D})-G[q_{\text{env}},q_{\text{in,B}},\pi_{B}]}{G[q_{\text{env}},q_{\text{in,B}},\pi_{B}]}
\end{equation}
so that if $s>s_{\text{max}}$ then $s$ is not achievable. We then consider the case where $0<s<s_{\text{max}}$. If $I_{B}>R(\Omega_{B})$, then genotype $B$ is suboptimal. In this case, genotype $A$ could adopt a sensory system and strategy that achieves $R(\Omega_{B})$ (down-pointing arrow in Fig. \ref{fig:rate-distortion_diagram}), requiring less information about the environment than genotype $B$ but achieving
\begin{equation}
s_{1}=\frac{\Gamma[I_{B}]-\Gamma\circ{}R(\Omega_{B})}{\Lambda^{\star}[q_{\text{env}},\delta_{y,c}]-\Omega_{B}-\Gamma[I_{B}]}>0.
\end{equation}
If $s<s_{1}$, then genotype $A$ can achieve $s$ with strictly less information than genotype $B$. If instead $s>s_{1}$, it may still be possible for genotype $A$ to achieve $s$ with less information than $B$, depending on the specific rate-distortion function and cost function. An iso-informational change is also possible (left-pointing arrow in Fig. \ref{fig:rate-distortion_diagram}), yielding a selection coefficient
\begin{equation}
s_{2}=\frac{\Omega_{B}-D(I_{B})}{\Lambda^{\star}[q_{\text{env}},\delta_{y,c}]-\Omega_{B}-\Gamma[I_{B}]}>0.
\end{equation}
This could be achieved by choosing the optimal sensing distribution $q_{\text{in}}(y|c)$ that yields $I_{B}$ and the corresponding optimal strategy. Of course, any distortion-rate pair above the rate-distortion curve is achievable, and the specific rate-distortion function and cost function are needed to fully answer the question at hand.

If instead genotype $B$ achieves a distortion-rate pair that falls on the rate-distortion curve ($I_{B}=R(\Omega_{B})$), then genotype $A$ can achieve a selection coefficient over genotype $B$
\begin{equation}
s_{3}=\frac{\Gamma\circ{}R(\Omega_{B})-\Gamma\circ{}R(\Omega_{A})+\Omega_{B}-\Omega_{A}}{\Lambda^{\star}[q_{\text{env}},\delta_{y,c}]-\Omega_{B}-\Gamma\circ{}R(\Omega_{B})}
\end{equation}
with distortion-rate pair $(\Omega_{A},R(\Omega_{A}))$. In order to achieve $0<s<s_{\text{max}}$, genotype $A$ can choose $\Omega_{A}$ such that $s=s_{3}$. The minimal amount of information gain for $A$ to achieve $s$ over $B$ is $R(\Omega_{A})-R(\Omega_{B})$. If $\Omega_{B}>\hat{D}$, a rate-distortion achieving genotype $A$ must achieve a smaller distortion than $B$ ($\Omega_{A}<\Omega_{B}$). Then, $R(\Omega_{A})-R(\Omega_{B})$ is positive and cost of the gain in information for $A$ over $B$ is offset by the decrease in distortion. If instead $\Omega_{B}<\hat{D}$, a rate-distortion achieving genotype $A$ must achieve a larger distortion than $B$ ($\Omega_{A}>\Omega_{B}$). In this case, the sensory system of $B$ is too expensive and $A$ can achieve a greater overall growth rate by reducing its sensory abilities, so that $R(\Omega_{A})-R(\Omega_{B})$ is negative. Finally, if $\Omega_{B}=\hat{D}$, the maximum selection coefficient achievable by $A$ is $s=0$.

\section{Conclusion}

This work introduces a formal analogy between fitness resulting from information and distortion. The rate-distortion function we have defined determines optimal trade-offs between growth and information. In the terminology of Shoval \emph{et al.} \cite{shoval2012evolutionary}, the rate-distortion function is a Pareto front in the tasks of maximally exploiting information about the environment and minimizing the metabolic costs of information processing. The rate-distortion framework captures the fact that sensory mechanisms are subject to evolutionary processes where previous work considered the accuracy of sensing as a given quantity. Given models of environmental sensing and population growth, one can calculate rate-distortion functions for real populations and examine whether the existing phenotypes fall near the rate-distortion curve or not. This type of analysis for a large number of populations would provide crucial empirical insight into the nature of information-cost trade-offs in evolution, and how these trade-offs differ across taxa and ecological contexts.

In general, it is difficult to find the explicit form of rate-distortion functions. For our modified function in Eq. \ref{eq:rate-distortion}, the task of numerically minimizing $I(C;Y)$ is further complicated by the additional optimization step required to calculate the mean distortion $\Omega^{\star}[q_{\text{env}},q_{\text{in}}]$ (Eq. \ref{eq:cost_of_imperfect_information}). Our method of approximating the mean distortion as a function of the distribution $q_{\text{in}}$ for use in constrained optimization of $I(C;Y)$ greatly simplifies the task of calculating the rate-distortion function. However, as the model of environmental sensing becomes more complicated this approach must be amended. For example, as the number of environmental states increases we must calculate a larger number of mean distortions in order for an approximation to be sufficiently accurate. In this case it may be useful to use methods such as Gaussian process regression, which has found success in a number of biological applications including estimating the fitness landscape of proteins \cite{romero2013navigating}. As discussed in Section \ref{sec:bilevel_opt}, approaches from bilevel optimization could also be of use in calculating the rate-distortion function.

The Kelly \cite{kelly1956new} framework we have used in this article is a convenient setting for theoretical analysis of information in evolution. However, its assumptions are overly restrictive, namely that growth is unlimited, that all variables are independent of their previous values, and that growth rates are frequency-independent. Fortunately, our formulation of the rate-distortion function is not dependent of the particular model of population growth, although there may not be a clear definition of the distortion function (Eq. \ref{eq:distortion_function}) in other models. The rate-distortion function in Eq. \ref{eq:rate-distortion} can be adapted to the particular model of growth or environmental sensing in use so that a more appropriate definition of mean distortion or of the accuracy of sensing may be substituted for $\Omega^{\star}[q_{\text{env}},q_{\text{in}}]$ or $I(C;Y)$.

\section{Acknowledgements}
This work was supported by the United States Defense Advanced Research Projects Agency RadioBio program under grant number HR001117C0125, and by a Discovery grant from the Natural Sciences and Engineering Research Council of Canada.

\appendix

\section{General methods}\label{appendix:methods}

We produced all figures using Matplotlib 3.1.1 \cite{hunter2007matplotlib} in a Python 3.7 Jupyter Notebook \cite{kluyver2016jupyter}. We performed calculations using SciPy \cite{2020SciPy-NMeth} (especially optimization) and NumPy \cite{harris2020array}.

\section{Proof of existence and uniqueness of a distortion $\hat{D}$ maximizing $G^{\star}(D)$}\label{appendix:gopt}

\begin{theorem}
Define $\mathcal{D}$ as the interval $\big[0,\Omega^{\star}[q_{\text{env}},\eta]\big]\subset\mathbb{R}$. Suppose that the rate-distortion function $R(D)$ is strictly convex and monotonically decreasing for $D\in\mathcal{D}$ and that the cost function $\Gamma[I]$ is (not necessarily strictly) convex and monotonically increasing for $I\in\mathbb{R}^{0+}$. Suppose that $R(D)$ and $\Gamma[I]$ are also both twice continuously differentiable on $\mathcal{D}$ and the image of $\mathcal{D}$ under $R$, respectively. Then there exists a unique $\hat{D}$ that maximizes $G^{\star}(D)$ in $\mathcal{D}$. 
\end{theorem}

\begin{proof}
Recall the definition of the overall growth rate
\begin{equation}
G^{\star}(D)=\Lambda^{\star}[q_{\text{env}},\delta_{y,c}]-D-\Gamma\circ{}R(D) 
\end{equation}
Because $\Gamma[I]$ and $R(D)$ are both twice differentiable, the overall growth function $G^{\star}(D)$ is twice differentiable and therefore continuous in $\mathcal{D}$. The set $\mathcal{D}$ is a closed, bounded subset of the real numbers, which is therefore compact. Using the Weierstrass theorem (see \cite{sundaram1996first}, Theorem 3.1), we can see that $G^{\star}(D)$ obtains a maximum $\hat{D}$ in $\mathcal{D}$, which is not necessarily unique.

To show that $\hat{D}$ is unique, we first prove that $G^{\star}(D)$ is strictly concave. Because both $\Gamma[I]$ and $R(D)$ are twice differentiable, we have
\begin{align}
\frac{d^{2}}{dD^{2}}\Gamma\circ{}R(D)=&~\bigg(\frac{d^{2}}{dR^{2}}\Gamma\circ{}R(D)\bigg)\bigg(\frac{d}{dD}R(D)\bigg)^{2}\\\nonumber
&+\bigg(\frac{d}{dR}\Gamma\circ{}R(D)\bigg)\bigg(\frac{d^{2}}{dD^{2}}R(D)\bigg)\\
>&~0
\end{align}
for all $D\in\mathcal{D}$. The first term is greater than or equal to zero due to the convexity of $\Gamma[I]$ while the second term is strictly greater than zero in $\mathcal{D}$ as $\Gamma[I]$ is monotonically increasing and $R(D)$ is strictly convex. Because $\frac{d^{2}}{dD^{2}}\Gamma\circ{}R(D)>0$, it is clear that $\Gamma\circ{}R(D)$ is strictly convex in the interior of $\mathcal{D}$ (see \cite{sundaram1996first}, Theorem 7.10). It is straightforward to show that $\Gamma\circ{}R(D)$ is strictly convex on all of $\mathcal{D}$ using the continuity of $\Gamma\circ{}R(D)$. Then, we can easily show that $G^{\star}(D)$ is strictly concave for $D\in\mathcal{D}$ from
\begin{equation}
\frac{d^{2}}{dD^{2}}G^{\star}(D)=-\frac{d^{2}}{dD^{2}}\Gamma\circ{}R(D)<0.
\end{equation}
The set $\mathcal{D}$ is convex and we have shown that $G^{\star}(D)$ is strictly concave on $\mathcal{D}$. As a consequence of these two facts, the optimal distortion $\hat{D}$ is unique (see \cite{sundaram1996first}, Theorem 7.14). The optimal distortion is defined as $\hat{D}$ satisfying
\begin{equation}\label{eq:critical}
\frac{d}{dD}G^{\star}(D)\Big|_{D=\hat{D}}=0
\end{equation}
if such a distortion exists within $\mathcal{D}$, or at one of the boundaries of $\mathcal{D}$, $\hat{D}=0$ or $\hat{D}=\Omega^{\star}[q_{\text{env}},\eta]$ otherwise.
\end{proof}

\bibliography{bibliography}

\end{document}